\documentclass[preprint,aps]{revtex4}
\usepackage{graphicx}%

\begin{document}
\title{Stock Market and Motion of a Variable Mass Spring}
\author{E. Canessa\footnote{E-mail: canessae@ictp.it}}
\affiliation{The Abdus Salam International Centre for Theoretical Physics,
Trieste, Italy
\vspace{3cm}
}

%%%%%%%%%%% Abstract
\begin{abstract}
We establish an 
analogy between the motion of spring whose mass increases 
linearly with time and volatile stock markets dynamics 
within an economic model based on simple temporal demand 
and supply functions [J. Phys. A: Math. Gen. 
{\bf 33}, 3637 (2000)]. The total system energy $E_t$ is shown to 
be proportional to a decreasing time dependent spring constant $k_t$. 
This model allows to derive log-periodicity $cos[log (t-t_{c})]$
on commodity prices and oscillations (surplus and shortages) in 
the {\it level of stocks}.  We also made an attempt to connect these
results to the Tsallis statistics parameter $q$ based on a possible
force-entropy correlation [Physica A {\bf 341}, 165 (2004)] and find 
that the Tsallis second entropic term $\sum_{i=1}^{W} p_i^{q}/(q-1)$ 
relates to the square of the demand (or supply) function.

\vspace{1.0cm}
\end{abstract}
\maketitle
%%%%%%%%%%%%%%%%%%%%

The interplay of demand and 
supply of reserves (and commodities) is primordial for the stability of the 
economy (and the market) in general. The latter should also relate to the 
existence of clear log-periodic periods of the form $cos[log (t-t_{c})]$ 
(or $sin[log (t-t_{c})]$) that characterize the daily dynamics of some of the 
most important indices worldwide \cite{Dro99,Bar05,Dro03}.
In this work we introduce a simple spring system gaining momentum 
from the surroundings in order to give further insight into the 
temporal log-periodic phenomenon.  We develop a 'toy' model
based on an analogy with mechanics that exhibits these
oscillations.  Furthermore an analogy with a 
standard economic model based on simple $t$-dependent demand and supply 
stock functions is presented \cite{Can00b}. We correlate this
result to the Tsallis statistics parameter $q$ based on a possible
force-entropy correlation \cite{Can04}.

The interest in log-periodicity is twofold. Firstly they enhance 
the fit quality to observed data with better precision than simple 
power laws by adjusting the (frequency, local minima and maxima of) 
oscillations. Secondly their real-time monitoring could, in principle, 
allow for an enhancement of predictions in different contexts
\cite{Sor02}. At the theoretical level, it has been argued that that 
there is a close relation between log-periodicity and the renormalization
group theory \cite{Sor96a,Can00a}. Logarithm periodic patterns also 
emerge from percolation models when applying the concept of clusters 
to groups of investors acting collectively \cite{Sta98,Sta00}. Besides, 
log-periodicities can be a consequence of transient clusters induced 
by an entropy-like term that may reflect the amount of cooperative 
information carried by the state of a large stochastic system of 
different species \cite{Can00}.  Alternatively,
a simple model was introduced in \cite{Sor02a}
based on the interplay between nonlinear positive feedback and reversal
in the inertia to account for accelerating log-periodic oscillations.
However, there is not yet a fundamental theory that can explain 
the universality of log-periodic motion on large time scales.  

The restoring force $F$ that a spring exerts on a fluctuating mass 
$m_t$ resting on a frictionless surface and attached to one end of 
the spring is defined by Hooke's law
\begin{equation}
F = {d (m_tv) \over dt} = 
  \left({dm_t \over dt}\right) v + m_t \left({dv \over dt}\right) \equiv  -kx \;\;\; ,
\end{equation}
where $x$ is the displacement from the spring equilibrium 
position $x_0$ at time $t_0$, $v = dx/dt$ is the system 
velocity and $k$ an spring constant.  This law holds for 
small changes in lenght.

The differential equation describing the spring's motion can then 
be approximated as
\begin{equation}\label{eq:motion}
\left({dm_t \over dx}\right) \left({dx \over dt}\right)^{2} +
   m_t \left({d^{2}x \over dt^{2}}\right) = -k_tx \;\;\; ,
\end{equation}
where, in general, the "spring constant" may also change
with time when momentum is gained from, or lost to, the 
surroundings ({\it e.g.}, rain or sand droping from a hopper 
at rest \cite{Hal78,Sou02,Dia03}).
Let us consider the simplest case in which the mass increases
linearly with time 
\begin{equation}\label{eq:mass}
m_t = m_0 \left({t \over t_0}\right) \;\;\; ,
\end{equation}
such that the slope $dm_t/dt \equiv m_0/t_0 > 0$ is the constant
rate at which mass is gain.  This could relate the asymptotic
behaviour of a brownian particle having a fluctuating mass \cite{Aus06}.
In other words our mass coefficient satisfies
\begin{equation}\label{eq:mass1}
{1 \over m_t}\left({dm_t \over dx}\right) = 
  {1 \over t}\left({dt \over dx}\right) = (vt)^{-1} \;\;\; . 
\end{equation}
The oppositive case in which the mass of the oscillator
decreases lineraly with time has been experimentally studied 
in \cite{Flo03}.

The addition of mass implies less resistance by a soft
sprint (with a smaller $k$) and thereafter a reduction of the spring 
constant as a function of time.  Let us assume then for simplicity
\begin{equation}\label{eq:spring}
k_t = k_0 \left({t_0 \over t}\right) \;\;\; .
\end{equation}
The time dependent angular frequency 
$\omega_t \equiv \sqrt{k_t/m_t}$,
which it has been shown to remain valid when the mass is variable
\cite{Flo03}, defines the {\it constant} angle
\begin{equation}\label{eq:omega}
\theta \equiv \omega_t \cdot t = \sqrt{{k_0 \over m_0}} \; t_0 
             = \omega_0 \cdot t_0 \equiv \theta_0 \;\;\; .
\end{equation}
Therefore the spring's equation of motion in eq.(\ref{eq:motion}) 
under these constraints reduces to
\begin{equation}\label{eq:motion1}
t^{2} \left({d^{2}x \over dt^{2}}\right) + 
  t \left({dx \over dt}\right) + \theta^{2} x = 0 \;\;\; .
\end{equation}
For $t \ge t_0 \ne 0$, the general solution has the form
\begin{equation}\label{eq:logp}
x(t) = x_0 \; sin\left[ \theta\; ln \left({t \over t_0}\right) \right] +  
 x_1 \; cos\left[ \theta\; ln \left({t \over t_0}\right) \right] \;\;\; ,
\end{equation}
with $x_{0,1}$ dimension constants.  It follows then that
the system in question displays temporal logarithm periodic oscillations 
as a consequence of adding mass linearly with time.

Setting $x_1=0$ in eq.(\ref{eq:logp}) for simplicity, the system velocity is found to 
display decreasing logarithm periodic oscillations such that 
\begin{equation}\label{eq:velocity}
v(t)=x_0 w_t \; cos\left[\theta\;
        ln\left({t \over t_0}\right)\right] = 
            {\theta \over t} \sqrt{x_0^{2}-x^{2}(t)} \;\;\; .
\end{equation}
The maximum and minimum peaks occur at the exponentially decaying time 
values $t_n = t_0 \; exp[-\pi/2\theta]$.  Hence the angle $\theta$ 
drives the peak positions of the oscillations.

Substitution of eq.(\ref{eq:velocity}) into (\ref{eq:mass1}) gives 
\begin{equation}\label{eq:mass2}
ln \; m_t = {1 \over \theta} \int
          {dx \over \sqrt{x_0^{2}-x^{2}(t)}} + const. 
\end{equation}
Since $sin \;\alpha = sin \; (2\pi n + \alpha)$, we then have
\begin{equation}\label{eq:mass3}
ln \left({m_t \over m_0}\right) = \left({1 \over \theta}\right)  
    sin^{-1}\; \left\{sin \left[\theta\; ln \left({t \over t_0}\right)\right] \right\} =
         ln \left({t \over t_0}\right) \;\;\; ,
\end{equation}
which means that $m$ grows up proportional to $t$ as initially
assumed in eq.(\ref{eq:mass}).

Let us establish next an analogy between the motion of a variable mass
spring in eq.(\ref{eq:motion1}) and the dynamics of stock markets 
within a simple 
economic model based on simple $t$-dependent demand $D(P)$ and supply 
$Q(P)$ functions for one commodity \cite{Can00,Can00a,Can00b}.
The market will be considered competitive so it
self-organizes to determine the behaviour of the asset price $P$
at time $t$ ({\it i.e.,} no individual producer can set his own
desired price). 

In a competitive market the price rate increases usually as a 
functional of the excess demand function $E(P)=D(P)-Q(P)$, 
such that $dP/dt \equiv f[ E(P) ]$ \cite{Cle84}.  
Since in general a commodity can be stored, then stocks of the commodity 
build up when the flow of output exceeds the flow of demand and 
vice-versa.  The rate at which {\it the level of stocks}, $S$, changes 
can then be approximated as $dS/dt = Q(P) - D(P)$.  Thus a price
adjustment relation that takes into account deviations of the stock
level $S$ above certain optimal level $S_{o}$ (to meet any demand 
reasonably quickly) is simply given by
\begin{equation}\label{eq:stock}
\frac{dP}{dt} = -\gamma \frac{dS}{dt} + \lambda (S_{o}-S)  \;\;\; ,
\end{equation}
where $\gamma$ ({\it i.e.}, the inverse of excess demand required
to move prices by one unity \cite{Bou98}) and $\lambda$ are positive
factors.  For $\lambda > 0$, prices increase when stock levels are 
low and raise when they are high (with respect to $S_{o}$). 
When $\lambda = 0$, the price adjusts at a rate proportional
to the rate at which stocks are either raising or running down. 

For all asset prices $P(t)$, simple forms for the  
demanded and supplied quantities are usually postulated
\begin{eqnarray}\label{eq:dq}
D(P) & = &  d^{*} + d_{o}(t) (P-P^{*}) \;\;\; ,
\nonumber \\
Q(P) & = &  q^{*} + q_{o}(t) (P-P^{*}) \;\;\; , 
\end{eqnarray}
where $d_{o}$, $q_{o}$ are temporal functions related to material costs,
wage rate, {\it etc} and $P^{*}=P(t^{*})$, $d^{*}=D(P^{*})$, 
$q^{*}=Q(P^{*})$ are values at equilibrium.  

To complete this simplest, economics model let us consider
as in \cite{Cle84} that $S_{o}$ depends linearly on the demand function
\begin{equation}\label{eq:stocklevel}
S_{o}(P) = \ell_{o} + \ell D(P) \;\;\; ; \;\;\; \frac{dS_{o}}{dt} = \ell \left(\frac{dD}{dt}\right)  \;\;\; ,
\end{equation}
with $\ell_{o}$ and $\ell$ constants. 
Therefore, in equilibrium (where $\frac{dP}{dt}|_{P^{*}}=0$ and
$\frac{dS}{dt}|_{S^{*}}=0$, so that demand equals supply 
$D(P^{*}) = Q(P^{*})$ and $S=S^{*}=S_{o}$),
we obtain $d^{*}-q^{*} = 0$ and $S^{*} = \ell_{o} + \ell d^{*}$.

After some algebra, it can be shown that the second derivative of 
the price variations in eq.(\ref{eq:stock}) then satisfies
\begin{equation}\label{eq:dyn}
\frac{d^{2}P}{dt^{2}} + [\;\gamma \beta_{o}(t) - \lambda\ell d_{o}(t)\;] \left(\frac{dP}{dt}\right) +
  [\;\gamma \left(\frac{dq_{o}}{dt}\right) - (\gamma + \lambda\ell) \left(\frac{dd_{o}}{dt}\right) + \lambda \beta_{o}(t)\;] 
                     (P-P^{*}) = 0 \;\;\; ,
\end{equation}
where $\beta_{o}(t) \equiv q_{o}(t)-d_{o}(t)$.
Since, in general, $d_{o} < 0$ and $q_{o} > 0$ then  
these conditions gives $\beta_{o}(t) > 0$ \cite{Can00b}.

From a comparison between the motion equation of oscillating systems 
that gain momentum from the surroundings as discussed in
eq.(\ref{eq:motion1}) and the price adjustment eq.(\ref{eq:dyn}) 
for one commodity, we readily identify 
\begin{eqnarray}\label{eq:analogy}
P(t)-P^{*} & \rightleftharpoons & x(t) \;\;\; , \nonumber \\ 
\gamma \beta_{o}(t) - \lambda\ell d_{o}(t) & \rightleftharpoons & \frac{1}{t}  \;\;\; , \nonumber \\ 
\gamma \left(\frac{dq_{o}}{dt}\right) - (\gamma + \lambda\ell) \left(\frac{dd_{o}}{dt}\right) + \lambda \beta_{o}(t) & \rightleftharpoons & \left(\frac{\theta}{t}\right)^{2} \;\;\; . 
\end{eqnarray}
This means to have analogous displacements, an analogous spring constant (related to
an analogous mass via the relation  $k_t m_t = k_0 m_0$) and an analogous frequency. 
The above also leads to these conditions for log-periodicity in
financial systems
\begin{eqnarray}\label{eq:analogy1}
\lambda\ell d_{o}(t) & \rightleftharpoons & \gamma \beta_{o}(t) - \frac{1}{t} \;\;\; , \nonumber \\ 
\lambda\ell q_{o}(t) & \rightleftharpoons & (\gamma + \lambda\ell) \beta_{o}(t) - \frac{1}{t} \;\;\; . 
\end{eqnarray}
Thus within our analogy, insight into the nature of the observed 
log-periodicy in stocks markets can be obtained from the
new perspective of the demand and supply functions. 
The expressions for $D$ and $Q$ of eq.(\ref{eq:dq}) in conjuction
with the above conditions for $d_{o}(t)<0$ and $q_{o}(t)>0$,
depict the fact that as price falls, the quantity demanded for a 
commodity can increase under temporal constrains
based in one of the basic principles of economy. 
That is, the higher the price, the higher the profit, then the higher
the supply.

For this class of systems we derive via eqs.(\ref{eq:logp}) and 
(\ref{eq:velocity}) and $x_1=0$, the universal relation 
\begin{equation}\label{eq:univ}
\left({x \over x_0}\right)^{2} + \left({v \over x_0 w_t}\right)^{2} 
   = 1 \;\;\; 
\end{equation}
for all time $t$, and the total system energy becomes
\begin{equation}\label{eq:energy}
E_t \equiv U + T = {1 \over 2}k_t x^{2} + {1 \over 2}m_t v^{2} = 
   {1 \over 2}k_t x_{0}^{2} =
        \left({k_t \over k_0}\right) U_0 \;\;\; ,
\end{equation}
where $U = \pm \int kx\;dx = \pm {1 \over 2}kx^{2}$ is the potential 
energy stored in the (compressed or elongated) spring and $T$ is the
kinetics energy. Hence $E_t$ is inversely proportional to $t$
(or proportional to $k_t$).
This general result can help to understand the reported log-periodic 
behaviour.  Since from eq.(\ref{eq:omega}) we have that $w_t$ is 
proportional to the inverse of time, then the total energy reaches 
a minimum only at $t \rightarrow \infty$.  Hence systems displaying 
log-periodic behaviour are far from equilibrium and unstable 
({i.e.}, volatility spanning over several months in the case of 
financial systems \cite{Van98}).
Let us conclude making an attempt to connect the variable mass spring model 
to Tsallis statistics \cite{Tsa88}.  This could be useful to gain new insights 
on the observed log periodicity in stock market  out of equilibrium (around times 
of financial turmoil) and the non-extensive entropic index $q$.  

We have recently introduced a statistical thermodynamic approach of moving 
particles forming an elastic body which leads to reveal molecular-mechanical 
properties of classical and nonextensive dynamical systems 
by assuming a simple multiplicative form for the probability 
$p_i=\mu_i\cdot\nu_i$ that the system is in 
the microstate $i(=1,\cdots,W)$, which satisfies the factorization 
$\sum_{i=1}^W p_i=\sum_{i=1}^N\mu_i\cdot\nu_i\equiv 1$ \cite{Can04}.
Expressions then follow for the  
the Helmholtz free energy $A=E-T{\cal S}$, considering the temperature $T$ and
volume $V$ to be independent variables, 
with $E$ the internal energy and ${\cal S}$ the entropy of the system
\begin{eqnarray}
A/k_{B}T & = &  \sum_{i=1}^{W} \mu_i\cdot\nu_i \; \ln \; \mu_{i} \;\;\; ,
                      \label{eq:helmholtz} \\
E/k_{B}T & = & - \sum_{i=1}^{W} \mu_i\cdot\nu_i \; \ln \;
               \nu_{i} \;\;\; ,  \\
  {\cal S}/k_{B} & = & -\; \sum_{i=1}^{W} \mu_i\cdot\nu_i \; \ln \; \mu_i\cdot\nu_i \;\;\; ,
               \label{eq:canessa}
\end{eqnarray}
with $k_{B}$ the Boltzman constant.
Since the moving particles are assumed to form an elastic body, then 
the system entropy can be correlated to a tensile force acting on the system
\begin{equation}\label{eq:newton}
{\cal F} =  \left(\; \frac{\partial A}{\partial x} \;\right)_{T} =
k_{B}T \; \frac{\partial}{\partial x} \left( \; \sum_{i=1}^{W} \mu_i\cdot\nu_i \; \ln \; \mu_{i} \; \right)_{T} =
 \left(\; \frac{\partial E}{\partial x} \;\right)_{T} - \;
 T \left(\; \frac{\partial {\cal S}}{\partial x} \;\right)_{T} \;\;\; .
\end{equation}
Tsallis statistics is derived assuming the thermal energy of the particles $k_{B}T$
to be proportional to their energy states $\epsilon$ by the nonextensivity
(integer) factor $q-1$ for all the $i$-microstates and $q \ne 1$. This was
shown to be equivalent to $\ln \; \nu_{i} = 1/1-q$.
Then it follows that
${\cal S}/k_{B}  \approx \left[1 - \sum_{i=1}^{W} (\mu_i\cdot\nu_i)^{q} \right] / q-1$,
which corresponds to the entropy term introduced by Tsallis and,
to a first approximation, a possible force-entropy correlation 
${\cal F}/k_{B}T \approx {\partial \over \partial x} \left( 
            {\sum_{i=1}^W p_i^{q} / q-1} \right)_{T}$.

A comparison of the latter and eq.(\ref{eq:energy})
for a system of variable mass springs gives 
\begin{equation}
- k_t x \rightleftharpoons k_{B}T {\partial \over \partial x} \left( 
            {\sum_{i=1}^W p_i^{q} \over q-1} \right)_{T} \;\;\; .
\end{equation}
This implies that the Tsallis second entropic term 
relates directly to the square of demand (or supply) function
given in eq.(\ref{eq:dq}) and using eqs.(\ref{eq:analogy}), namely
\begin{equation}
- \; {1 \over 2} \left[ {k_0 t_0 \over d_{o}(t)} [\gamma \beta_{o}(t) - \lambda\ell d_{o}(t)]
        (D(P)-d^{*})\right]^{2}  \rightleftharpoons k_{B}T \left( 
            {\sum_{i=1}^W p_i^{q} \over q-1} \right)_{T} \;\;\; .
\end{equation}
This relation also provides some 
insight into possible temporal dynamics of the non-extensive systems
and gives physical significance to the q-values.

In this way stock market turmoil (in terms 
of the behaviour of the demand and supply functions) belong to a class of oscillating 
systems that gain momentum from the surroundings and, in turn, may correlate to
a class of non-extensive dynamical systems described by Tsallis statistics.
We have considered both a variable mass and variable "spring constant".  
This interplay seems to be a necessary and sufficient condition to understand 
log-periodicity on commodity prices and oscillations like surplus and shortages in
the {\it level of stocks}, relating prices adjustments, as for example in crude 
oil and petroleum products.  Since, we have shown
\begin{equation}
{dS \over dt } = Q(P) - D(P) \; \propto \; P(t) - P^{*} \; \propto \; x(t) 
       \; \propto \; cos[log (t-t_{c})] \;\;\; ,
\end{equation}
then it is plausible that policy makers like the European Central Bank \cite{ECB08} could 
exploit the present results to handle demand/supply for (oil) reserves by looking at log-periodic
oscillations \cite{EIA09}.  When (oil) prices are high it is expensive to keep surplus of stock,
whereas in the case of low prices it makes sense to accumulate stocks for future gains.
This balance of the supply and demand defines about 80\% of the resulting price (the
difference is defined by speculation) \cite{Vyg02}.

\end{document}